\documentclass[intlimits,twoside,a4paper]{article}
\usepackage{amsmath,amssymb}
\usepackage{graphicx}

\usepackage[T2A]{fontenc}
\usepackage[cp1251]{inputenc}

\usepackage[eqsecnum]{cmpj2}

\usepackage[english,ukrainian]{babel}

\usepackage{lscape}
\usepackage{url}
\usepackage{color}

\newcommand{\ukr}[1]{\selectlanguage{ukrainian}#1\selectlanguage{english}}

\issue{2012}{15}{4}{47101}

\doinumber{10.5488/CMP.15.47101}

\articletype{Chronicle}

\title[Marian Smoluchowski:  A story behind one photograph]%
{Marian Smoluchowski:  A story behind one  photograph}
\author[A. Ilnytska \textsl{et al.}]
{A. Ilnytska\refaddr{label1}, \ J. Ilnytskyi\refaddr{label2},
Yu. Holovatch\refaddr{label2}, A. Trokhymchuk\refaddr{label2}}
\addresses{
\addr{label1} V. Stefanyk Lviv National Scientific Library of Ukraine, 79000 Lviv, Ukraine
\addr{label2} Institute for Condensed Matter Physics of the National
Academy of Sciences of Ukraine, \\ 1 Svientsitskii St., 79011 Lviv, Ukraine
}

\date{Received October 29, 2012, in final form December 3, 2012}
\authorcopyright{A. Ilnytska, \ J. Ilnytskyi, Yu. Holovatch, A. Trokhymchuk, 2012}

\begin{document}
\selectlanguage{english}
\maketitle
\begin{abstract}
We discuss the photograph procured from the archives of the V.Stefanyk Lviv
National Scientific Library of Ukraine dated by 1904 which shows
Marian Smoluchowski together with professors and graduate students of
the Philosophy department of the Lviv University.
The personalia includes both the professors and the graduates
depicted on the photograph with the emphasis on the graduates as being
much less known and studied.
The photograph
originates from the collection of the Shevchenko Scientific Society, therefore a brief historical background on the activities of physicists in this society around
that period of time is provided as well.
\keywords  history of science, Shevchenko Scientific Society, Lviv University, Marian Smoluchowski
\pacs 01.60.+q, 01.65.+g
\end{abstract}

\section{About the photograph and its origin}

\begin{landscape}
\begin{figure}[!h]
\centerline{\includegraphics[width=20cm]{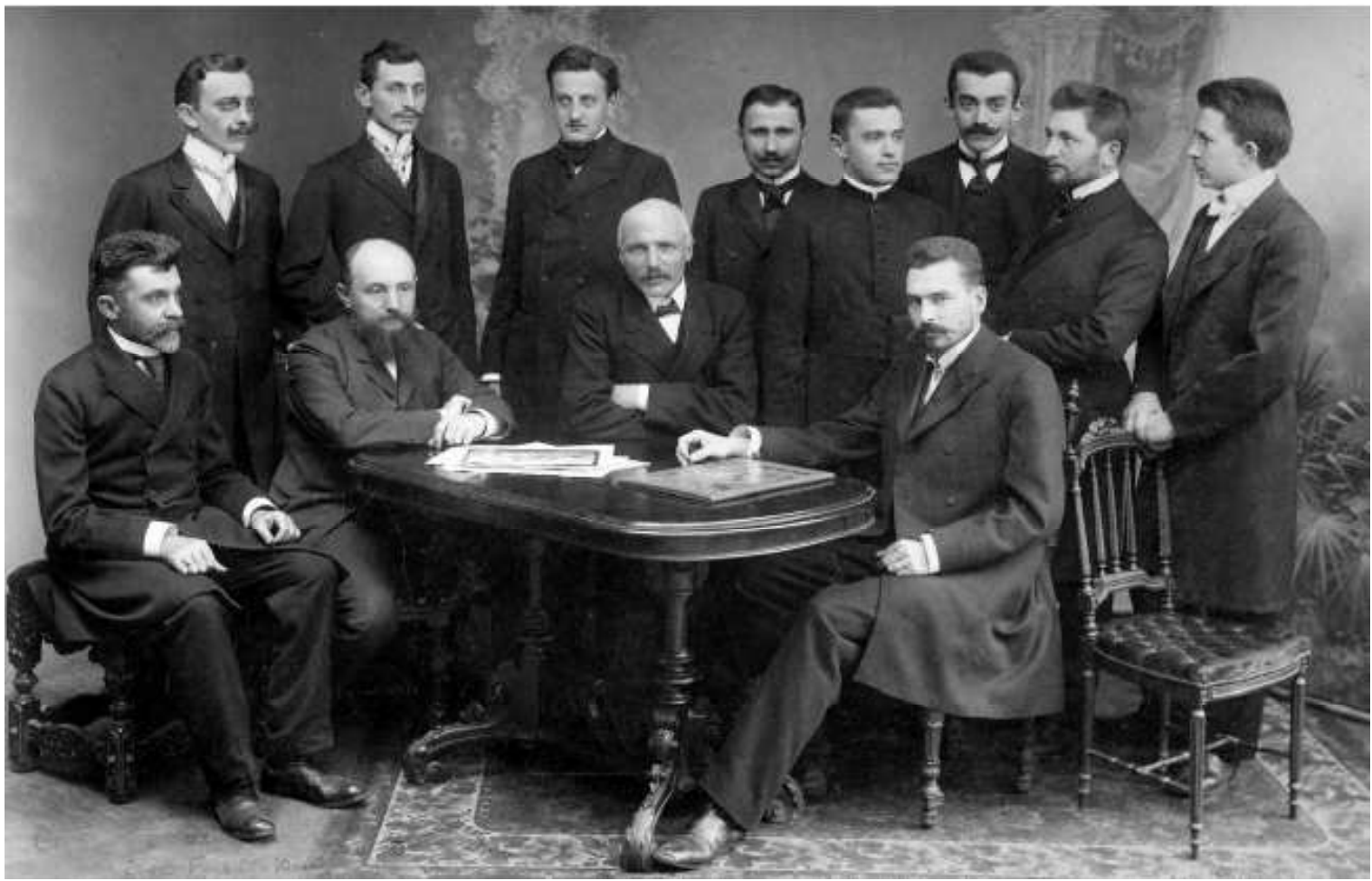}}           
\vspace*{8pt}
\caption{\label{fig1}{\bf Marian Smoluchowski together with professors and 4-th year
    graduates of the Lviv University (June, 1904)}. Seated (left to right):
    Prof.~Jan Rajewski, Prof.~Ignacy Zakrzewski, Prof.~J\'ozef Puzyna,
    Prof.~Marian Smoluchowski. Standing (left to right): J\'{o}zef
    Bartkiewicz, Ivan Bodnar, Antoni {\L}omnicki, Nykyfor Danysh,
    Rev. Volodymyr Ardan, Zdzislaw Thullie, Roman Shypaylo,
    Ivan Sitnytsky. {\em From the archives and with the permission from
    the V.~Stefanyk Lviv National Scientific Library of Ukraine.}}
\end{figure}
\end{landscape}

Not so many photographs of Marian Smoluchowski are known. To our
best knowledge, the photographs representing Marian Smoluchowski together with his
colleagues and students are altogether absent. That is why the
photograph (see figure~\ref{fig1}) recently discovered in the
archives of the Institute for Library Art Resources Studies which
belongs to the V.~Stefanyk Lviv National Scientific Library of Ukraine
seems to be especially valuable.  The picture on the photograph has
dimensions $22.8\times29.7$~cm and shows four professors of the
Philosophy department of Lviv University and eight 4th year
absolvents (graduates). All this information is the part of the
handwritten description that is found on the frontside of the
photograph and includes the date (June, 1904) and all the
names written in Ukrainian. The {backside} contains the year
(1904) and the names written by ink either in Polish or in Ukrainian (possibly at the date the photograph was taken).

The photograph was found among a collection that once belonged to the
museum of Shevchenko Scientific Society (abbreviated hereafter as NTSh~--- from its name in Ukrainian: \ukr{{\em Наукове Товариство
ім. Шевченка}})~\cite{enc,NTSh}.  The NTSh museum was closed in 1940 and,
as a result, part of its collection was transferred to the newly
established Academic Library in Lviv (now the V.~Stefanyk Lviv National
Scientific Library of Ukraine). Currently all these materials are
stored in the archives of the Institute for Library Art Resources
Studies that is  part of this library.

The collection of the NTSh museum, including the photographs, was
filled-up either by the purchases made by the museum or owing to the
gifts obtained from the patrons. New additions were reported
systematically in the {\em Chronicle of Shevchenko Scientific Society}
in the section {\em Museum Status}. For instance, in the report for
the years 1923--1925, the then manager of the museum Yuriy Polyansky
wrote that ``the previous years brought further progress into the
development of the museum. New exhibits started to be collected and the existing collections enlarged\ldots{}
At the end one needs to note that
our citizens started to be interested in the museum fortune. This is
evident from quite a number of gifts that arrived at the museum and from
the increase of the number of visitors''~\cite{Polan_1926}. How exactly
the photograph, which is the
\noindent subject of our study, became a part of
the collection of the NTSh museum is not known.  One of the
possibilities is that it might have belonged to Ivan Bodnar, one of the
graduates depicted on it, who was president of NTSh during the years
1940--1944, and could have well offered it to the museum.

The NTSh society was founded in Lviv in December 1873. It was named
after famous Ukraine's national poet and artist Taras
Shevchenko. After being reorganized in 1892, the society acquired features
of a typical scientific association and served as the first unofficial
Ukrainian Academy of Arts and Sciences. One of its sections was
devoted to mathematics, natural sciences and medicine.  We will
provide here some brief historical account on the development of physics within NTSh
society and on the physicists who were members of NTSh.

Various fields of physics were represented in the NTSh society during
different time periods of its activity\footnote{Shevchenko Scientific
Society was liquidated by Soviet power in 1940, in 1947 it renewed its
activity in Western Europe and in the USA, in 1989 it was restored in
Ukraine~\cite{Holovatch92a,Holovatch92b}.}. Among the first full NTSh members
(academicians) who were elected in 1899, one finds the names of Ivan
Puluj, Volodymyr Levytsky and Petro Ohonovsky.  The first of them,
Ivan Puluj (1845--1918) is known for his work in the fields of
physics, electrical engineering and theology\footnote{Together with
P.~Kulish and I. Nechuy-Levytsky he made the first translation of the Old
and the New Testament  from Ancient Greek to Ukrainian
\cite{Gajda01,Pului_works}.}. The major fields of Puluj's
interests in physics were molecular physics, cathode rays and X-rays.
In particular, by using the electric discharge tubes of his own
construction, Puluj obtained fundamental results on cathode rays.
Within a variety of vacuum lamps in electrical engineering there is one known
as a 'Puluj lamp'.  Early in 1896, a month and half after
R\"ontgen's publication on the discovery of X-rays, Ivan Puluj has
submitted two papers~\cite{Pului_pap_a,Pului_pap_b} where he provided a deep explanation of the origin
of X-rays and discovered their ionization properties. There are many evidences that the Puluj tube gave the highest quality X-ray
pictures \cite{Gajda01}. Volodymyr Levytsky (1872--1956) for a long
time was the head of the mathematical-natural sciences-medical section within NTSh
and was the NTSh president in 1931--1935. Although his principal
field of activity was mathematics, he also contributed to physics
by publishing papers, by compiling Ukrainian physical
dictionaries, and by writing textbooks on physics in Ukrainian~\cite{Levytsky12}.
Another NTSh full member, Petro Ohonovsky (1853--1917) is known as
the author of the first physics textbook in Ukrainian~\cite{Ohonovsky97}.

Later on among the full NTSh members one can find more representatives
of different fields of physics and engineering. In particular, these
are: Yulian Hirniak (1881--1970)~--- physical chemistry; Roman
Tseheljsky (1882--1956)~--- physics of electrolytes; Volodymyr Kucher
(1885--1959)~--- quantum mechanics and quantum statistics; Stepan
Tymoshenko (1878--1972)~--- mechanics; Ivan Feshchenko-Chopivsky (1884--1952)~--- metallurgy;
 Oleksander Smakula (1900--1983)~--- optics, known
in particular, owing to the discovery of anti-reflective coating of
lenses; Andrij Lastovetsky (1902--1943)~--- spectroscopy; Zenon
Khraplyvy (1904--1983)~--- relativistic quantum mechanics. In 1920-ies
Max Plank (1858--1947),
Albert Einstein (1879--1955) and Abram Ioffe (1880--1960)  were elected  full members of the NTSh society.  Many
interesting documents about contacts of these and other scientists
with the NTSh society are now available in the NTSh archives, main parts of
which are stored at the Central State Historical Archive of Ukraine in
Lviv~\cite{CDIA}, V.~Stefanyk Lviv National Scientific Library of
Ukraine, National Library of Warsaw~\cite{Svarnyk05} and some other
institutions\footnote{See reference~\cite{Svarnyk05} for a comprehensive
description of main sources.}.

\section{Personalia}

The photograph shown in figure~\ref{fig1} reveals four mature professors
sitting in the front, and eight handsome 4th year graduates in their
youth (all around 24 years of age) standing behind them.  The
professors, from left to right, are Jan Rajewski, Ignacy Zakrzewski,
J\'ozef Puzyna and Marian Smoluchowski, all of whom were lecturing physics and
mathematics in the year the photo was taken. Since the Lviv period of
{\bf Marian Smoluchowski's}
life is a subject of a separate paper in this issue~\cite{Rovenchak12},
we will omit
detailed information on Smoluchowski here, referring the readers to
the above mentioned sources as well as to references~\cite{cmp12,epj13,smol_about_a,smol_about_b,smol_about_c}.
The
short information given below concerns other three professors~\cite{Duda07,Holovatch06,Enc11}.

{\bf Jan Rajewski} (1857--1906) was born in Lesser
Poland\footnote{Malopolska (Polish), which is one of the historical
regions of Poland, its capital is the city of Krak\'ow.}. He attended
gymnasiums (high schools) in Drohobych and Lviv.  During 1875--1879 Jan
Rajewski studied mathematics and physics in the Philosophy
department of the Lviv University and in 1884 he received here a doctoral
degree. In the 1883--1890 Rajewski worked as a school
teacher of mathematics and physics in Lviv, Stanislav and Krak\'ow.
In 1900 Rajewski became an extraordinary professor of mathematics
at Lviv University.  His works are devoted to the theory of
differential equations.  He is buried in Lviv at Lychakiv Cemetery.

{\bf Ignacy Zakrzewski} (1860--1932) was born in Ternopil and graduated from
Lviv University in 1882. He worked at Lviv University (1882--1886), at Krak\'{o}w University (1887--1891), and studied at Berlin
University (1892). In 1893 Ignacy Zakrzewski became a professor.
During 1892--1920 he was head of the chair of experimental
physics at Lviv University.
Main scientific interests include physics of ice and
temperature behavior of the heat capacity of solids.

{\bf J\'ozef Puzyna} (1856--1919) was born in Novy Martyniv (province
of Stanislav\footnote{Ivano-Frankivsk at present.}).  In 1875 he
finished the gymnasium in Lviv and studied mathematics at Lviv
University. After defending a doctoral thesis in 1883, Puzyna continued
his studies at Berlin University, where one of his teachers was Karl
Weierstrass. Since 1885 Puzyna lectured at Lviv University and in 1892 he
became an ordinary professor and headed the chair of mathematics.
During 1905 (the next year after the photograph in figure~\ref{fig1} was taken)
J\'ozef Puzyna was rector, and in 1906~--- vice rector of Lviv
University. Puzyna's main field of activity was the theory of analytic
functions. Among his students there were Volodymyr Levytsky, Hugo Steinhaus,
Antoni {\L}omnicki, Waclaw Sierpi\.{n}ski.

Obviously, all professors shown in the photograph were already well
established in their career whereas the destiny of the eight young men
after graduation from the Lviv University could provide a high-quality
literary plot (in a way German romanticist E.T.A. Hoffmann was
inspired by the engravings by Jacques Callot in his {\em
Fantasiest\"{u}cke in Callots Manier}). We intentionally concentrated
on the life stories of the graduates, as far as these are
much less known
as compared to that of their teachers.

The first standing from the left is {\bf J\'{o}zef Bartkiewicz} and to
our regret we found nothing on his biography data.  The second
standing from the left is {\bf Ivan Bodnar} \ukr{(Іван Боднар)}. The
following information can be found in the Encyclopedia of Modern
Ukraine~\cite{Bodnar_encycl}. Ivan Bodnar was born on 1~September,
1880 in the village Vovkiv (now Peremyshlyany district of Lviv
region).  From 1900 Bodnar attended Vienna University. In
1905 he graduated from the Philosophy department of Lviv
University. Then, Ivan Bodnar taught mathematics and physics at
Ternopil Teachers Training Seminary (1906--1921) and later on at both Peremyshl
Gymnasium and Ternopil Gymnasium (1921--1931). In 1930th he worked at the
Ternopil Ukrainian Bank, and in 1940--1941 and 1944--1950 at the Academic
Library in Lviv (now V.~Stefanyk Lviv National Scientific Library of
Ukraine).  During 1941--1944 Ivan Bodnar worked as an
accountant at the Lviv City administration.

More sketches to the personality of Ivan Bodnar can be found in the
historical studies, especially on his involvement in politics in 1918~\cite{Bodnar_Lazarovych} and his role in the social and cultural life
of Ternopil in 1920--1930~\cite{Bodnar_Ostapiuk,Ternopil_site}. In
particular, Ivan Bodnar is mentioned as head of the Teachers' Society
and co-founder of the Ladies Institute in Ternopil~\cite{Ternopil_site}.
Among others, Ivan Bodnar was a subject of
coordinated night searches performed by authorities in September
1930.

Since 1910 Ivan Bodnar was a member of the NTSh. During 1939--1940 he
was vice-president and in 1940--1946 he was president of the
NTSh.  Professor Ivan Bodnar died on 18~December, 1968 in Lviv~\cite{Bodnar_encycl}.

According to the indication on the front side of the photograph, the
third graduate from the left is {\bf Antoni {\L}omnicki}.  Although
the handwriting on the back side is not so clear (reads more like
``Hordynski''), we are quite sure that this must be {\L}omnicki\footnote{We compare the image in figure~\ref{fig1} with the later photographs of
professor Antoni {\L}omnicki.}.

Antoni {\L}omnicki was born on 17~January, 1881 in Lviv.  He studied
at the Lviv University and the University of G\"{o}ttingen, attending
courses given by H.~Minkowski, D.~Hilbert and F.~Klein\footnote{D.~Hilbert and F.~ Klein were members of the NTSh.}.
In 1920 {\L}omnicki became a professor of Lviv Polytechnic and his
field of interests included mathematical analysis, probabilistics,
statistics, cartography, didactics, applied mathematics.  In 1938
{\L}omnicki became a member of the Warsaw Scientific Society (TWN)~\cite{Lomnicki_enc}.  In 1920--1922, one of the founders of modern
functional analysis Stefan Banach was an assistant
of Prof.~{\L}omnicki. Later {\L}omnicki was a major adviser for
Banach's doctoral thesis until
his habilitation at Lviv University.

Antoni {\L}omnicki tragically died on 4~July, 1941 in Wzg\'{o}rza
Wuleckie (part of Lviv) in the massacre of Lviv professors by Nazi
occupation forces~\cite{Lviv_mass}.  In December 1944 Stefan Banach
wrote the following tribute to professor {\L}omnicki: ``A native of
Lviv, he worked for over twenty years as a mathematics professor at
Lviv Polytechnic.  He prepared hundreds of engineers for their
profession. I was his assistant. He was the first to instil in me the
importance and responsibility of a professor's task. He was an
unrivaled educator, one of the best I have ever known. He was the author of
many popular schoolbooks as well as textbooks on advanced analysis for
technologists, surpassing in quality those published abroad. His work
in the field of cartography was of a high level. Equally effective
were his efforts as an instructor and pedagogue.  Professor {\L}omnicki had
tremendous energy and a great work ethic''~\cite{Lomnicki_wiki}.

The fourth graduate from the left is {\bf Nykyfor Danysh}
\ukr{(Никифор Даниш)}. He was born in 1877.  After graduating from Lviv
University Nykyfor Danysh dedicated himself to teaching, was active as
an organizer of various private schools in Kolomyia and Stanislav.
In particular, in 1908 Nykyfor Danysh started teaching at
Kolomyia Ukrainian Gymnasium, alongside with Roman Shypaylo, another
graduate depicted on the photograph that we are discussing~\cite{Danysh_Kolom_hist}. Danysh wrote several books on
the history of the Kolomyia Gymnasium~\cite{Danysh_Kolom_hist,Danysh_book1}.  In
turmoil of the end of 1918, Nykyfor Danysh was a member of the of the West
Ukrainian People's Republic (ZUNR) administration
in Chortkiv (Ternopil region)~\cite{Danysh_ZUNR}.

In 1921, being a member of the Ukrainian Teachers Society, Nykyfor
Danyzh established the Ukrainian Ladies Gymnasium in Stanislav and
became its first director. The memoirs of a former student of this
gymnasium gave him an account of being a great teacher and the true
patriot~\cite{Danysh_Lemekha}. Nykyfor Danysh can be found in the
photograph of all graduates of Kolomyia Gymnasium during the years
1914--1939 (see photo in reference~\cite{Danysh_later_photo}, where Nykyfor
Danysh is eighth from the left sitting in the first row). Nykyfor
Danyzh immigrated to United States where he died in 1954~\cite{Danysh_UW}.

The fifth graduate from the left is {\bf Reverend Volodymyr Ardan}
\ukr{(о. Володимир Ардан)}. He was born in Polyany, Korosnyansky
povit (Lemkivshchyna region) and was  a professor of mathematics
in the State Ukrainian Gymnasium in
Peremyshl~\cite{Ardan_Shakh_MIzh_Syanom}.
In 1905 he took an active position establishing a pre-gymnasium
hostel for pupils coming from the nearby villages and until 1910
was a director of this hostel~\cite{Ardan_Shakh_Sribnolenty_Syan}.

In 1934 the Roman Throne established the Lemky Apostle Administration
and Rev. V.~Ardan became a member of the Lemky Capitula in Rymanov,
holding a position of synod judge~\cite{Ardan_Shakh_MIzh_Syanom}.
In his memoirs Rev.~V.~Hrynyk  also recalls Volodymyr Ardan as a great
mathematician and teacher in the Peremyshl Gymnasium and mentions
that he was arrested during the Nazi occupation for unclear reasons~\cite{Ardan_Hrynyk}.

The sixth graduate from the left is {\bf Zdzislaw Thullie}.  He was a
son of a well known scientist, Prof. Maksymilian Thullie (1853--1939),
twice rector of the Lviv Politechnic, expert in bridge
construction, senator and an active Christian-Democrat politician.
Zdzislaw Thullie was interested in the theory of metals. We found his
paper (in Polish) published in 1908~\cite{Thullie_PMF,Thullie_PMF_2} and a book (in
French) published in 1912~\cite{Thullie_book}. He  was also a professor
and taught physics at the first Lviv Real School.

Zdzislaw Thullie was in the middle of his career and was to be
habilitated soon when his life was abruptly terminated by a tragic
accident that happened on 16~June, 1922.  The story (based on contemporary
newspapers) is retold by Stanis{\l}aw S. Nicieja~\cite{Thullie_death}. Z.~Thullie together with two other professors
({\L}opusza\'{n}ski and Zag\'{o}rski) took a large group of their
students for a trip to Bubnyshcha (rocky region near Bolekhiv, now
Ivano-Frankivsk district) to have some leisure before the exams. The
sunny day was spent among the rocks, and the night~--- inside the
tents.  On the next day, after breakfast, all walked back towards the
train station to return to Lviv in time for the teachers conference
(in which the professors participated). However, around noon, a sudden
storm with lightning broke up. Z.~Thullie with two students separated
from the rest and stopped by the river to watch for rough waves.
Professor, in romantic mood, said to pupils: ``Watch, guys, what a
wonderful and at the same terrifying view!''. Next moment the
lightning tore the sky with the roar and Prof.~Thullie fell down
struck by it. None of immediate first aid measures could help as the
death was instantaneous, and the most that students could do was to bring
the body to Lviv for the funeral which took place a few days later and
was attended by many.

The seventh graduate from the left is {\bf Roman Shypaylo} \ukr{(Роман
Шипайло)}. After graduation, since 1906 he worked as a teacher in
Kolomyia Ukrainian Gymnasium, alongside with Nykyfor Danysh~\cite{Danysh_Kolom_hist}. Besides teaching mathematics and physics,
Roman Shypaylo was also known as a great enthusiast of Ukrainian choir
music. He organized and directed the school choir which performed much
of the available contemporary Ukranian repertoire (among others, such
large-scale compositions as {\em Kavkaz} by S.~Ludkevych and {\em Haydamaky}
by J.~Kyshakevych)~\cite{Shypajlo_choir}.  In the newspapers of that
time one can find the articles by R.~Shypaylo~\cite{Shypajlo_choir2}.
Mary Beck (1908--2005), the first woman elected to Detroit's City Council, who studied in
1920th in Kolomyia, characterized Shypajlo as a former soldier (he
served in the Ukrainian Galician Army, UHA), having a romantic soul and as
a great admirer of photography~\cite{Shypajlo_Beck}. Roman Szypajlo
also initiated and patronized the 13th girls Plast (Ukrainian scout)
regiment.

The rightmost graduate depicted in the photograph is {\bf Ivan
Sitnytsky} \ukr{(Іван Сітницький)}. He was born on 25~June, 1881 in
Komarno (the village near Lviv) and died on 10~October, 1947 in Lviv~\cite{Sitn_EU}. Ivan Sitnytsky taught exact sciences at Lviv
Academic Gymnasium~\cite{Sitn_Gymn}. He published several textbooks on
geometry, chemistry and physics. Sitnytsky was a member of the NTSh.
Similarly to Roman Shypaylo, Sitnytsky also served as a centurion in
the UHA.  Early in 1919 the Artillery School was
established in Stryi, then it was moved to Stanislav.
The school was
well equipped for teaching and supplied with a good amount of cannons and
horses. After its first commandant officer was moved to another
regiment, his place was taken by Ivan Sitnytsky~\cite{Sitn_Tkachuk}.

The jubileum book for the Lviv Academic Gymnasium~\cite{Sitn_Jubil_book} contains numerous references and memoirs
concerning Prof.~I.~Sitnytsky. It is mentioned that he taught
mathematics in higher classes introducing Ukrainian terminology. The
lectures on physics were also given by him
(reference~\cite{Sitn_Jubil_book}, p.~467). Together with Dr. Chaikivsky
they lectured the seminar {\em Problems and methods in Physics}
in the {\em Circle for Natural sciences} (reference~\cite{Sitn_Jubil_book},
p.~179).

\section{Instead of summary}

Marian Smoluchowski (1872--1917) was an outstanding scientist, one of
the founders of the modern statistical physics. His life was bright
but, unfortunately, very short. Native Polish, Marian Smoluchowski was
born in Vienna and lived there until receiving a doctoral degree.  At
the age of forty five Marian Smoluchowski unexpectedly died in
Krak\'ow, shortly after moving there from Lviv, where he spent his
fourteen most fruitful years in science. Although there are
numerous studies~\cite{smol_about_a,smol_about_b,smol_about_c,Rovenchak12} dedicated to the life
and scientific activities of Marian Smoluchowski, some questions still
remain open. One of them concerns the pupils and/or coworkers that
Smoluchowski could have in Lviv.  The present sketches do not answer
these questions directly, but still shed some light regarding the
scientific (and, to some degree, political) atmosphere at those times.
We hope that both the photograph and our sketchy study, could be
helpful for
 those interested in the history of science and physics in particular.

\section*{Acknowledgement}

We are thankful to Yaroslav Hrytsak, Roman Plyatsko and Halyna Svarnyk for useful suggestions.


\ukrainianpart

\title{Маріан Смолуховський: історія одного фото}
\author{А.~Ільницька\refaddr{label1}, Я.~Ільницький\refaddr{label2},
  Ю.~Головач\refaddr{label2}, А.~Трохимчук\refaddr{label2}}
\addresses{
\addr{label1} Львівська національна наукова бібліотека України
ім. В.~Стефаника, 79000 Львів, Україна\
\addr{label2} Інститут фізики конденсованих систем НАН України,
вул. Свєнціцького, 1, 79011 Львів, Україна}

\makeukrtitle

\begin{abstract}
\tolerance=3000

Розглянуто фотографію датовану 1904 роком з Архіву Львівської
національної наукової бібліотеки ім. В.~Стефаника, на якій зображено
професорів та випускників четвертого року навчання філософського
факультету Львівського університету. Світлина походить із колекції
Наукового Товариства ім. Шевченка, тому подано коротку історичну
довідку про діяльність цього товариства на межі ХIX-XX століть.
Персоналія включає коротку довідку про зображених на фото професорів
та ширшу інформацію про випускників (як таку, яка менш доступна в
довідкових джерелах).

\keywords історія науки, Наукове Товариство ім. Шевченка, Львівський
університет, Маріан Смолуховський
\end{abstract}
\end{document}